\documentclass{article}
\usepackage{spconf,amsmath,graphicx}
\usepackage{amsfonts}
\usepackage{mathtools}
\usepackage[english]{babel}

\title{Temporal Order Preserved Optimal Transport-based Cross-modal Knowledge Transfer Learning for ASR}
\name{Xugang Lu$^{1}$, Peng Shen$^{1}$, Yu Tsao$^{2}$, Hisashi Kawai$^1$}
\address{1. National Institute of Information and Communications Technology, Japan\\
	2. Research Center for Information Technology Innovation, Academia Sinica, Taiwan}
\begin{document}
%\ninept
%
\maketitle
\begin{abstract}
Transferring linguistic knowledge from a pretrained language model (PLM) to an acoustic model has been shown to greatly improve the performance of automatic speech recognition (ASR). However, due to the heterogeneous feature distributions in cross-modalities, designing an effective model for feature alignment and knowledge transfer between linguistic and acoustic sequences remains a challenging task. Optimal transport (OT), which efficiently measures probability distribution discrepancies, holds great potential for aligning and transferring knowledge between acoustic and linguistic modalities. Nonetheless, the original OT treats acoustic and linguistic feature sequences as two unordered sets in alignment and neglects temporal order information during OT coupling estimation. Consequently, a time-consuming pretraining stage is required to learn a good alignment between the acoustic and linguistic representations. In this paper, we propose a Temporal Order Preserved OT (TOT)-based Cross-modal Alignment and Knowledge Transfer (CAKT) (TOT-CAKT) for ASR. In the TOT-CAKT, local neighboring frames of acoustic sequences are smoothly mapped to neighboring regions of linguistic sequences, preserving their temporal order relationship in feature alignment and matching. With the TOT-CAKT model framework, we conduct Mandarin ASR experiments with a pretrained Chinese PLM for linguistic knowledge transfer. Our results demonstrate that the proposed TOT-CAKT significantly improves ASR performance compared to several state-of-the-art models employing linguistic knowledge transfer, and addresses the weaknesses of the original OT-based method in sequential feature alignment for ASR.
\end{abstract}
\begin{keywords}
Optimal transport, Cross-modal knowledge transfer, automatic speech recognition
\end{keywords}
\section{Introduction}
The combination of a pretrained language model (PLM) with an end-to-end (E2E)-based acoustical model for automatic speech recognition (ASR) has made significant progress in recent years \cite{Li2022,Chan2016, Kim2017, Hori2017,Watanabe2017,RNNTASR}. The advantage of incorporating a PLM in ASR lies in the availability of unpaired large text corpora for training the PLM. Moreover, the linguistic knowledge encoded in the PLM can be utilized in ASR decoding. In most studies, the PLM is employed as an external language model (LM) for post-processing tasks such as beam search or rescoring in ASR \cite{BERTScore,MLMScore}. However, using an external LM for post-processing compromises the speed and sometimes parallel decoding capabilities of ASR. Addressing how to transfer linguistic knowledge to acoustic encoding during model training, and subsequently conducting speech recognition without relying on any external LM post-training, is an intriguing research topic. 

In this study, our focus is on transferring linguistic knowledge from a PLM to a temporal connectionist temporal classification (CTC)-based ASR \cite{CTCASR}. While there are several advanced end-to-end (E2E)-based ASR approaches that incorporate linguistic knowledge in acoustic model learning \cite{HierarchicalCTC,intermediateCTC}, using a PLM, such as bidirectional encoder representation from transformers (BERT) \cite{BERT}), facilitates linguistic knowledge transfer in ASR \cite{FNAR-BERT,NARBERT,KuboICASSP2022,Choi2022}, This knowledge transfer can also occur with a pretrained acoustic encoder, such as wav2vec2 \cite{wav2vec2.0}, for both linguistic and acoustic knowledge transfer \cite{Futami2022,Higuchi2023,CIFBERT1, wav2vecBERTSLT2022,CTCBERT1,CTCBERT2,DengICASSP2024,CIFBERT2}. However, due to the heterogeneous feature distributions in acoustic and linguistic spaces, it remains a challenging task to efficiently align feature representations between linguistic and acoustic modalities to facilitate knowledge transfer. In most studies, a cross-attention module is designed to integrate acoustic and text representations within a transformer decoder framework for combining acoustic and linguistic knowledge in ASR \cite{Transformer}. Yet, in the decoding stage, true text representations are unavailable, leading to the adoption of predicted text representations in decoding. This mismatch between training and testing phases weakens the benefits of linguistic information in ASR. 

For efficient alignment and matching, an effective distance metric is needed to measure the difference between acoustic and linguistic feature representations. Considering this requirement, optimal transport (OT) emerges as a suitable tool for cross-modal alignment and linguistic knowledge transfer. OT, originally proposed for optimal allocating resources  and later as a measure of discrepancies between probability distributions \cite{VillanoBook,CourtyNIPS2017}, has found widespread applications in machine learning, particularly in domain adaptation \cite{CourtyNIPS2017}. In the field of speech, it has been employed for cross-domain spoken language recognition and speech enhancement \cite{LuICASSP2021,Lin2021,ICLR2023}, as well as in speech translation and understanding \cite{Cho2020,Cross2021,ACL2023,ICML2023}. The OT has been initially proposed for linguistic knowledge transfer learning for ASR in \cite{ASRU2023Lu}. While OT is applicable for cross-domain alignment, its use in speech encounters a limitation. The original OT treats acoustic and linguistic feature sequences as two unordered sets in alignment and neglects temporal order information during OT coupling estimation. While acoustic and text speech exhibit a strong temporal order structure, requiring preservation of their temporal order relationship during alignment between an acoustic sequence and a linguistic sequence. Therefore, in \cite{ASRU2023Lu}, a well pretrained acoustic model was applied in order to efficiently explore matched acoustic features to those linguistic features during cross-modal learning. And the performance was strongly depended on the goodness of the pretrained acoustic model. In this paper, we propose a Temporal Order Preserved OT (TOT)-based Cross-modal Alignment and Knowledge Transfer (CAKT) model (TOT-CAKT) for CTC-based ASR. With the TOT-CAKT, the temporal order relationship is explicitly maintained during feature alignment and matching. It is hypothesized that through this alignment and matching, linguistic knowledge can be efficiently transferred to acoustic encoding, thereby enhancing ASR performance. 

The rest of this paper is organized as follows: the proposed method is introduced in Section \ref{sec:proposed}, where a cross-modal alignment module based on TOT and a neural adapter module for efficient linguistic feature transfer are designed. In Section \ref{sec:exp}, we conduct experiments to evaluate TOT-CAKT, comparing the results with several knowledge transfer learning algorithms for ASR, and provide a visualization of the learned transport coupling in OT. Finally, the conclusion is presented in Section \ref{sec:conclusion}.   
%\vspace{-2mm}
\section{Proposed method}
\label{sec:proposed}
\begin{figure}[tb]
	\centering
	\includegraphics[width=7cm, height=4.5cm]{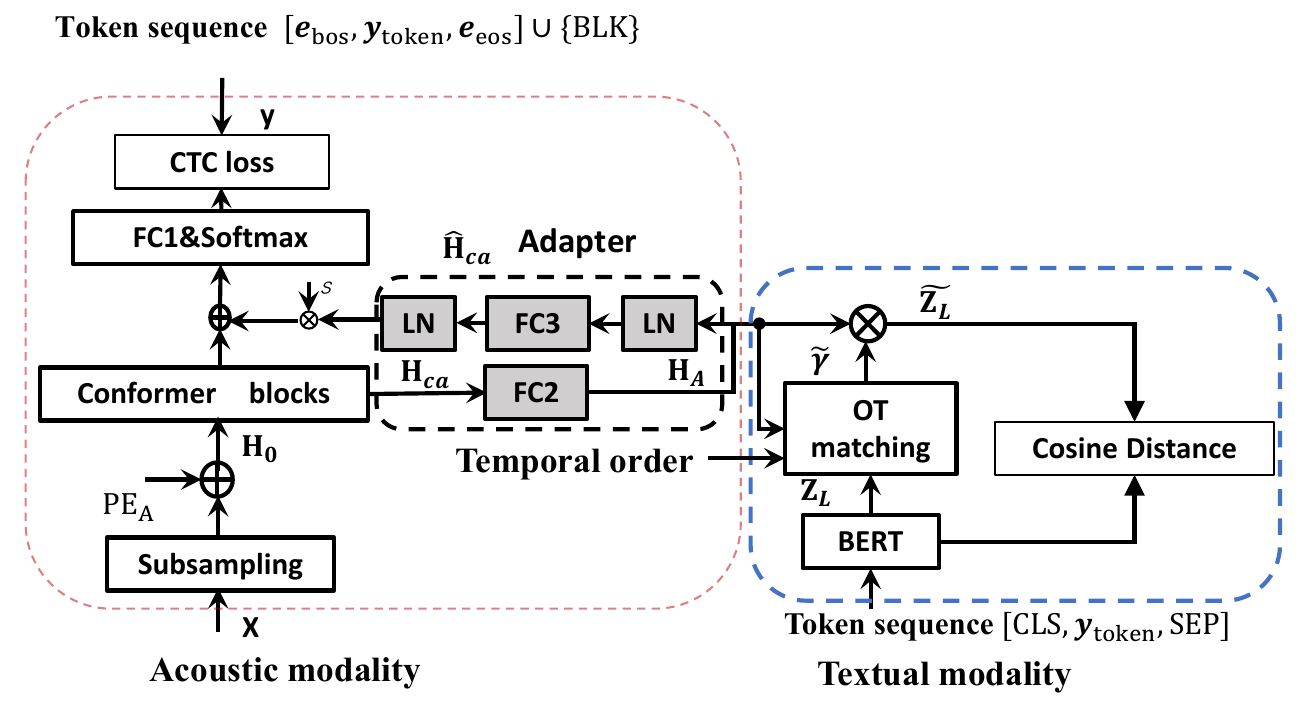}
	\caption{The model framework of the proposed cross-modal alignment and knowledge transfer method based on TOT.}
	%\vspace{-4mm}		
	\label{fig:frame1}
\end{figure}

The model framework of the proposed TOT-CAKT method is illustrated in Fig. \ref{fig:frame1}. This model framework is modified based on a conformer-CTC-based ASR model, incorporating two key modifications. First, an `Adapter' module is added as shown in the gray blocks in Fig. \ref{fig:frame1}. Second, an temporal order preserved OT-based cross-modal matching module is introduced in the right branch of Fig. \ref{fig:frame1}. Both the acoustic features extracted from the conformer encoder and linguistic features derived from a PLM (with BERT utilized in this paper) are involved in the cross-modal matching process. Further details are provided in the following sections.
\subsection{Acoustic and linguistic feature representations}
The acoustic feature is extracted from the acoustic encoder where a conformer-based encoder \cite{conformer2020}) is adopted. The process in the `Subsampling' module involves a two-layer convolution process with a downsampling operation (a downsampling rate of $4$ was used in this paper). By incorporating a positional encoding from ${\rm PE}_{\rm A}$, the initial input to conformer blocks is obtained as ${\bf H}_0 $. The output of the conformer encoder is represented as an acoustic representation ${\bf H}_{\rm ca}$.
\begin{equation}
	\begin{array}{l}
		{\bf H}_0  = {\rm Subsampling}\left( {\bf X} \right) + {\rm PE}_{\rm A}  \\ 
		{\bf H}_{{\rm ca}}  = {\rm Conformer}\left( {{\bf H}_0 } \right) \in R^{l_a  \times d_a },  \\ 
	\end{array}
	\label{eq:conformer}
\end{equation}
where $l_a$ and $d_a$ are temporal length and dimension of the acoustic feature vectors, respectively. Before engaging in cross-modal feature alignment, a linear projection termed `FC2' in the `Adapter' module is utilized to perform a feature dimension matching transform:
\begin{equation}
	{\bf H}_{\rm A}  = {\rm FC}_{\rm 2} \left( {{\bf H}_{{\rm ca}} } \right) \in R^{l_a  \times d_t } 
	\label{eq:FC2}
\end{equation}
In this equation, $d_t$ corresponds to linguistic feature dimension. In the right branch of Fig. \ref{fig:frame1}, the context-dependent linguistic feature representation is explored from a pretrained BERT model. The process is formulated as: 
\begin{equation}
	\begin{array}{l}
		%\begin{align}					
		{\bf y}_{{\rm token}}  = {\rm BERTTokenizer}\left( {\bf y} \right) \\ 
		\vspace{2mm}
		{\bf Z}_0  = \left[ {{\rm CLS, }{\bf y}_{{\rm token}} ,{\rm SEP}} \right] \\ 
		\vspace{2mm}
		{\bf Z}_i  = {\rm BERT}_i \left( {{\bf Z}_{i - 1} } \right), \\ 		
		%\end{align}
	\end{array}
	\label{eq:bert}
	%\vspace{-1mm}
\end{equation}
where `${\rm BERT}_{i}$' is the $i$-th transformer encoder layer of BERT model, $i$ takes values from $1$ to $L$, with $L$ representing the total number of BERT encoder layers. `$\rm BERTTokenizer$' is a process to convert standard text to word piece based tokens \cite{BERT}. Token symbols `CLS' and `SEP' represent the start and end of an input sequence. ${\bf Z}_L  \in \mathbb{R}^{l_t  \times d_t } $ is the final text representation which encodes context dependent linguistic information, $l_t$ denotes the sequence length, and $d_t$ represents feature dimension of text encoding representation. 
\subsection{Sinkhorn algorithm for cross-modal alignment}
The original OT was formulated to transform from one probability distribution to another with minimum transport cost \cite{VillanoBook}. In this study, we applied OT for feature alignment on two sets. Given acoustic and linguistic feature sequences ${\bf H}_A$ and ${\bf Z}_L$ respectively, as: 
\begin{equation}
	\begin{array}{l}
		{\bf H}_A  = \left[ {{\bf h}_1 ,{\bf h}_2 ,...,{\bf h}_i ,...,{\bf h}_{l_a } } \right] \\ 
		{\bf Z}_L  = \left[ {{\bf z}_1 ,{\bf z}_2 ,...,{\bf z}_j ,...,{\bf z}_{l_t } } \right], \\ 
	\end{array}
	\label{eq:twoseqs}
	%\vspace{-1mm}
\end{equation}
where $l_a$ and $l_t$ are lengths of the two sequences. Suppose the two sequences in Eq. (\ref{eq:twoseqs}) are sampled from two probability distributions with weight vectors ${\bf a}  = \left[ {a_1 ,a_2 ,...,a_i ,...,a_{l_a } } \right]$ and 
${\bf b}  = \left[ {b_1 ,b_2 ,...,b_j ,...,b_{l_t } } \right]$. ($a_i  = {1 \mathord{\left/
		{\vphantom {1 {l_a }}} \right.
		\kern-\nulldelimiterspace} {l_a }},b_j  = {1 \mathord{\left/
		{\vphantom {1 {l_t }}} \right.
		\kern-\nulldelimiterspace} {l_t }}$ as uniform distributions if no prior information is available). The OT distance between the two sequences is defined as:       
\begin{equation}
	L_{\rm OT} \mathop  = \limits^\Delta  \mathop {\min }\limits_{\gamma  \in \prod {\left( {{\bf H}_A ,{\bf Z}_L } \right)} } \left\langle {\gamma ,{\bf C}} \right\rangle, 
	\label{eq:ot}
\end{equation}
where ${\gamma}$ is a transport coupling set defined as:
\begin{equation}
	\prod {\left( {{\bf H}_A ,{\bf Z}_L } \right)} \mathop  = \limits^\Delta  \left\{ {\gamma  \in R_ + ^{l_a  \times l_t } \left| {\gamma {\bf 1}_{l_t }  = {\bf a},\gamma ^T {\bf 1}_{l_a }  = {\bf b}} \right.} \right\}
	\label{eq:coupling}
\end{equation}
In Eq. (\ref{eq:coupling}), ${\bf 1}_{l_a }$ and ${\bf 1}_{l_t }$ are vectors of ones with dimensions $l_a$ and $l_t$, respectively. In Eq. (\ref{eq:ot}), $\bf C$ is a distance matrix (or ground metric) with element ${c_{i,j} }$ defined as pair-wised cosine distance:
\begin{equation}
	c_{i,j}  = {\bf C} \left( {{\bf h}_i ,{\bf z}_j } \right)\mathop  = \limits^\Delta  1 - \cos \left( {{\bf h}_i ,{\bf z}_j } \right)
	\label{eq:Cost}
	%\vspace{-1mm}
\end{equation}
A fast estimation of OT has been introduced through the celebrated entropy-regularized OT (EOT) \cite{Cuturi2013} where the EOT loss is defined as:
\begin{equation}
	L_{\rm EOT} \left( {{\bf H}_A ,{\bf Z}_L } \right)\mathop  = \limits^\Delta  \mathop {\min }\limits_{\gamma  \in \prod {\left( {{\bf H}_A ,{\bf Z}_L } \right)} } \left\langle {\gamma ,{\bf C}} \right\rangle  - \alpha _1   H\left( \gamma  \right), 
	\label{eq:EOTloss}  
\end{equation}
where $\alpha _1 $ is a regularization coefficient, and $H\left( \gamma  \right) $ is entropy of coupling matrix defined as:
\begin{equation}
H\left( \gamma  \right) \mathop  =  - \sum\limits_{i,j} {\gamma _{i,j} \log \gamma _{i,j} } .	
\end{equation}
The solution of Eq. (\ref{eq:EOTloss}) can be implemented with Sinkhorn algorithm as \cite{VillanoBook}:
\begin{equation}
	\gamma _{\alpha _1 }  = diag\left( {{\bf u}_1 } \right)*{\bf G}*diag\left( {{\bf u}_2 } \right)
	\label{eq:gamma}
\end{equation}
where ${\bf G} = \exp \left( { - \frac{{\bf C}}{{\alpha _1 }}} \right)$, ${{\bf u}_1 }$ and ${{\bf u}_2 }$ are two scaling (or re-normalization) vectors.
\subsection{Temporal order preserved OT}
In the original estimation of OT in Eq. (\ref{eq:EOTloss}), the two sequences in Eq. (\ref{eq:twoseqs}) are treated as two sets without considering their temporal order relationship. In speech, temporal order information is crucial in OT coupling during cross-modal alignment, meaning that neighboring frames in an acoustic sequence should be progressively coupled with the neighboring tokens in a linguistic sequence. Therefore, as showed in Fig. \ref{fig:frame1}, the temporal order information is input to the OT matching block. For the sake of clarity, the two sequences in Eq. (\ref{eq:twoseqs}) can be further represented with temporal order information as:
\begin{equation}
	\begin{array}{l}
		{\bf H}_A  = \left[ {\left( {{\bf h}_1 ,1} \right),\left( {{\bf h}_2 ,2} \right),...,\left( {{\bf h}_i ,i} \right),...,\left( {{\bf h}_{l_a } ,l_a } \right)} \right] \\ 
		{\bf Z}_L  = \left[ {\left( {{\bf z}_1 ,1} \right),\left( {{\bf z}_2 ,2} \right),...,\left( {{\bf z}_j ,j} \right),...,\left( {{\bf z}_{l_t } ,l_t } \right)} \right] \\ 
	\end{array}	
	\label{eq:inputseq}
\end{equation} 
During the alignment of the two sequences for knowledge transfer, it is crucial to consider that elements with significant cross temporal distances might not be likely to be coupled. In other words, the coupling pairs with high probabilities between the two sequences should be distributed along the diagonal line of the temporal coherence positions. Based on this consideration, the temporal coupling prior could be defined as a two dimensional Gaussian distribution \cite{Su2017}. The fundamental concept is that the coupled pairs should not deviate significantly from the diagonal line of temporal coherence positions between the two sequences, which can be defined as:    
\begin{equation}
	p_{i,j} \mathop  = \limits^\Delta  \frac{1}{{\sigma \sqrt {2\pi } }}\exp \left( { - \frac{{d_{i,j}^2 }}{{2\sigma ^2 }}} \right),
	\label{eq:P}
\end{equation}
where $\sigma$ is a variation variable controlling the impact of the cross-temporal distance $d_{i,j}$ as defined in Eq. (\ref{eq:dij}).
\begin{equation}
	d_{i,j}  = \frac{{\left| {{\raise0.5ex\hbox{$\scriptstyle i$}
					\kern-0.1em/\kern-0.15em
					\lower0.25ex\hbox{$\scriptstyle {l_a }$}} - {\raise0.5ex\hbox{$\scriptstyle j$}
					\kern-0.1em/\kern-0.15em
					\lower0.25ex\hbox{$\scriptstyle {l_t }$}}} \right|}}{{\sqrt {{\raise0.5ex\hbox{$\scriptstyle 1$}
					\kern-0.1em/\kern-0.15em
					\lower0.25ex\hbox{$\scriptstyle {l_a^2 }$}} + {\raise0.5ex\hbox{$\scriptstyle 1$}
					\kern-0.1em/\kern-0.15em
					\lower0.25ex\hbox{$\scriptstyle {l_t^2 }$}}} }}
	\label{eq:dij}
\end{equation}
In Eq. (\ref{eq:dij}), the cross-temporal distance is defined on the normalized sequence lengths in acoustic and linguistic spaces. In this definition, it is evident that the farther the distance between a paired position and the temporal diagonal line, the lower possibility of their correspondence in transport coupling. By incorporating this temporal coherence prior as regularization, the new OT is defined as:
\begin{equation}
	L_{\rm{TOT}} ({\bf H}_A ,{\bf Z}_L )\mathop  = \limits^\Delta  \mathop {\min }\limits_{\mathclap{\gamma  \in \prod {({\bf H}_A ,{\bf Z}_L )} }}  < \gamma ,{\bf C} >  - \alpha _1   H(\gamma ) + \alpha _2   KL(\gamma ||P),
	\label{eq:TOP}
\end{equation}
where $\alpha _1$ and $\alpha _2$ are two trade off parameters. In Eq. (\ref{eq:TOP}), $KL(\gamma||P)$ is the Kullback-Leibler (KL) divergence between the transport coupling matrix $\gamma$ and temporal prior correspondence matrix $P$ with elements defined in Eq. (\ref{eq:P}). Building upon the definitions of KL-divergence and entropy, Eq. (\ref{eq:TOP}) can be further expressed to:
\begin{equation}
	L_{\rm{TOT}} ({\bf H}_A ,{\bf Z}_L )\mathop  = \limits^\Delta  \mathop {\min }\limits_{\gamma  \in \prod {({\bf H}_A ,{\bf Z}_L )} }  < \gamma ,{\bf \tilde C} >  - \tilde \alpha H(\gamma ),
	\label{eq:TOPC}
\end{equation}
where $\tilde \alpha  = \alpha _1    + \alpha _2$, and combined ground cost matrix as
\begin{equation}
	{\bf \tilde C} = {\bf C} - \alpha _2  \log P
	\label{eq:newC}
\end{equation}
Following the procedures outlined in \cite{Cuturi2013}, the solution of Eq. (\ref{eq:TOPC}) is obtained using the Sinkhorn algorithm as:
\begin{equation}
	\gamma _{\tilde \alpha }  = diag\left( {{\bf u}_1 } \right)*{\bf \tilde G}*diag\left( {{\bf u}_2 } \right),
\end{equation}
where ${\bf \tilde G} = \exp \left( { - \frac{{{\bf \tilde C}}}{{\tilde \alpha }}} \right)$. Substituting variables in Eq. (\ref{eq:newC}) to ${\bf \tilde G}$, we can obtain:
\begin{equation}
	{\bf \tilde G} = P^{\frac{{\alpha _2}}{{\alpha_1    + \alpha_2  }}} \exp ( - \frac{{\bf C}}{{\alpha_1  + \alpha_2 }})	
	\label{eq:newG}	
\end{equation}
From this equation, we can see that the transport coupling between the two sequences is further constrained by their temporal order correspondence. TOT involves several hyper-parameters that can be challenging to control. For the sake of simplification, we consolidate their effects into a reduced number of hyper-parameters. For example, considering Eq. (\ref{eq:newC}), the impact of variation $\sigma$ in Eq. (\ref{eq:P}) and $\alpha_2$ in Eq. (\ref{eq:TOP}) can be combined into a single control parameter $\beta$, defined as:
\begin{equation}
	{\bf \tilde C} = {\bf C} + \beta d_{i,j}^2 
	\label{eq:combined}
\end{equation}
And the Sinkhorn algorithm is applied on the cost function matrix ${\bf \tilde C}$ for OT in real implementations.  
\subsection{Loss function}
The proposed TOT-CAKT involves two loss functions: the cross-modal alignment and matching loss (in the right branch of Fig. \ref{fig:frame1}) and the CTC loss (in the left branch of Fig. \ref{fig:frame1}). In cross-modal alignment, the acoustic feature can be projected onto the linguistic space using OT as:
\begin{equation}
	\begin{array}{l}
		\begin{aligned}
			{\bf \tilde Z}_L &\mathop  = \limits^\Delta  {\rm OT}\left( {{\bf H}_{A}  \to {\bf Z}_{L} } \right) \\ 
			&= \gamma ^*  \times {\bf H}_{A}  \in R^{l_t  \times d_t },  \\ 
		\end{aligned}
	\end{array}	
\end{equation}
where $\gamma ^*$ is the optimal transport coupling based on OT. Subsequently, the alignment loss is defined as:
\begin{equation}
	L_{{\rm align}}  = \sum\limits_{j = 2}^{l_t  - 1} {1 - \cos \left( {{\bf \tilde z}_L^j ,{\bf z}_L^j } \right)}, 
	\label{eq:Align}
\end{equation}
where ${\bf \tilde z}_L^j$ and ${\bf z}_L^j$ are row vectors of feature matrices ${\bf \tilde Z}_{L}$ and ${\bf Z}_{L}$ (matching on temporal dimensions), respectively. In Eq. (\ref{eq:Align}), the usage of indices from $2$ to $l_t -1$ is for handling special symbols `$\rm CLS$' and `$\rm SEP$'. For efficient linguistic knowledge transfer to acoustic encoding, the following transforms are designed as indicated in Fig. \ref{fig:frame1}:
\begin{equation}
	\begin{array}{l}
		{\bf \hat H}_{{\rm ca}}  = {\rm FC}_{\rm 3} \left( {{\rm LN(}{\bf H}_{A} {\rm )}} \right) \in R^{l_a  \times d_a }  \\ 
		{\bf H}^{a,t}  = {\bf H}_{{\rm ca}} {\rm  + s \cdot LN(}{\bf \hat H}_{{\rm ca}} {\rm )}, \\ 
	\end{array}
	\label{eq:adapter}
\end{equation}
where $s$ is a scaling parameter to adjust the importance of transferring linguistic projected feature. Based on this new representation ${\bf H}^{a,t}$ which is intended to encode both acoustic and linguistic information, the final probability prediction for ASR is formulated as:
\begin{equation}
	{\bf \tilde P} = {\rm Softmax}\left( {{\rm FC1}\left({\bf H}^{a,t} \right)} \right),
	\label{eq:softmaxadd}
\end{equation}
where `FC1' is a linear full-connected transform. The total loss in model learning is defined as:
\begin{equation}
	L\mathop  = \limits^\Delta  \lambda .L_{{\rm CTC}} ({\bf \tilde P},{\bf y}_{{\rm token}} ) + (1 - \lambda ).w.{(L_{{\rm align}}  + L_{{\rm TOT}} )},   
	\label{eq:totalloss} 
\end{equation}
where $L_{{\rm CTC}} ({\bf \tilde P},{\bf y}_{{\rm token}} )$ is CTC loss, ${L_{{\rm align}} }$ and ${L_{{\rm TOT}} }$ are cross-modality alignment loss and TOT loss, respectively. After the model is trained, only the left branch of Fig. \ref{fig:frame1} is retained for ASR inference.  

\section{Experiments}
\label{sec:exp}
ASR experiments were conducted on the open-source Mandarin speech corpus AISHELL-1 \cite{AISHELL1} to evaluate the proposed algorithm. The data corpus comprises three datasets: a training set with 340 speakers (150 hours), a development (or validation) set with 40 speakers (10 hours), and a test set with 20 speakers (5 hours). Data augmentation as used in \cite{AISHELL1} was applied. Given the tonal nature of the Mandarin language in the ASR task, in addition to using 80-dimensional log Mel-filter bank features, three extra acoustic features related to fundamental frequency, i.e.,  F0, delta F0 and delta delta F0, were utilized as raw input features. These features were extracted with a 25ms window size and a 10ms shift. 
\subsection{Model architecture}
In Fig. \ref{fig:frame1}, the `Subsampling' module consists of two CNN blocks with 256 channels, kernel size 3, stride 2, and ReLU activation function in each. The acoustic encoder is formed by stacking $16$ conformer blocks \cite{conformer2020}, with each having a kernel size of 15, attention dimension $d_a=256$, $4$ attention heads, and a 2048-dimensional FFN layer. The `bert-base-chinese' from huggingface is used as the pretrained PLM for linguistic knowledge transfer \cite{Huggingface}. In this Chinese BERT model, $12$ transformer encoders are applied, the token (or vocabulary) size is 21128, and the dimension of linguistic feature representation is $d_t=768$. 
\subsection{Hyper-parameters in model learning}
Several hyper-parameters are associated with the proposed model, and these parameters may have a joint (or correlated) effect in efficient linguistic knowledge transfer learning. In our preliminary experiments, for easy implementation, they were fixed as $\beta=0.5$ in Eq. (\ref{eq:combined}), alignment trade off parameter $\lambda=0.3$ and scale parameter $w=1.0$ in Eq. (\ref{eq:totalloss}). ${\tilde \alpha }$ in Eq. (\ref{eq:TOPC}) and $s$ in Eq. (\ref{eq:adapter}) were varied in experiments. For optimization, Adam optimizer \cite{Adam} is used with a learning rate (initially set to 0.001) schedule with 20,000 warm-up steps. The model with cross-modal transfer was trained for 130 epochs, and the final model used for evaluation was obtained by averaging models from the last 10 epochs (the original conformer-CTC model without the adapter module was pretrained or trained from scratch in different experimental settings when jointly combined with cross-modal learning). The performance was evaluated based on character error rate (CER).   

\subsection{Results}
\label{sec:results}
In inference stage, only the left branch (blocks in dashed red box in Fig. \ref{fig:frame1}) is utilized, maintaining the decoding speed similar to that of the CTC-based decoding. In our experiments, only CTC greedy search-based decoding was employed, and the results are presented in table \ref{tab1}. The results of the baseline system and several state-of-the-art systems that integrate BERT for linguistic knowledge transfer are also provided for comparison. 
\begin{table}[tb]
	\centering
	\caption{ASR performance on the AISHELL-1 corpus, CER (\%).}
	\begin{tabular}{|c||c||c|}
		\hline
		Methods &dev set &test set\\
		\hline
		Conformer+CTC (Baseline)  &5.53 &6.05 \\
		\hline	
		\hline	
		Conformer+CTC/AED (\cite{Watanabe2017,wenet2.0})  &4.61 &5.06 \\						
		\hline
		NAR-BERT-ASR (\cite{NARBERT}) &4.90 &5.50 \\
		\hline
		KT-RL-ATT (\cite{CTCBERT1}) &4.38 &4.73 \\
		\hline
		Wav2vec-BERT (\cite{wav2vecBERTSLT2022}) &4.10 &4.39 \\
		\hline
		\hline
		\multicolumn{3}{|c|}{Without pretraining condition} \\
		\hline
		OT-BERT(w/o) (\cite{ASRU2023Lu}) &4.21 &4.52 \\
		\hline
		\textbf{TOT-CAKT}(w/o),$\tilde \alpha=0.01$,$s=1.0$ &4.03 &4.30 \\
		\hline
		\textbf{TOT-CAKT}(w/o), $\tilde \alpha=0.1$,$s=1.0$ &4.16 &4.48 \\
		\hline
		\textbf{TOT-CAKT}(w/o),$\tilde \alpha=0.5$,$s=1.0$ &3.99 &4.36 \\
		\hline
		\textbf{TOT-CAKT}(w/o),$\tilde \alpha=0.5$,$s=0.5$ &4.06 &4.40 \\	
		\hline
		\textbf{TOT-CAKT}(w/o),$\tilde \alpha=0.5$,$s=0.1$ &\textbf{3.93} &4.35 \\	
		\hline
		\textbf{TOT-CAKT}(w/o),$\tilde \alpha=0.01$,$s=0.1$ &4.02 &\textbf{4.29} \\		
		\hline
		\hline		
		\multicolumn{3}{|c|}{With pretraining condition} \\
		\hline		
		OT-BERT(w/)(\cite{ASRU2023Lu}) &3.96 &4.27 \\
		\hline
		\textbf{TOT-CAKT}(w/) &\textbf{3.88} &\textbf{4.21} \\								
		\hline
	\end{tabular}
	%\vspace{-3mm}
	\label{tab1}
\end{table}
In this table, `Conformer+CTC' is the baseline system, trained without linguistic knowledge transfer. `Conformer+CTC/AED' denotes a hybrid CTC/AED ASR system \cite{Kim2017, Hori2017,Watanabe2017} which used a transformer decoder with attention to text representation during model training. `NAR-BERT-ASR', `KT-RL-ATT', and `Wav2vec-BERT' are all based on integrating acoustic and linguistic features from BERT for ASR \cite{CTCBERT1,CTCBERT2,NARBERT,wav2vecBERTSLT2022}, and even used a pretrained acoustic model (from wav2vec2.0 \cite{wav2vec2.0}) and PLM for knowledge transfer. In the OT based cross-modal learning, two experimental conditions were examined. One is that models with cross-modal learning were trained from scratch, i.e., without pretraining condition. The other is that the models were initialized with a pretrained acoustic model, then were further trained with cross-modal learning, i.e., with pretraining condition. Correspondingly, in table \ref{tab1}, `OT-BERT(w/o)' and `TOT-CAKT(w/o)' denote results of cross-modal linguistic knowledge transfer learning based on OT in \cite{ASRU2023Lu} and the proposed TOT-CAKT method for without pretraining condition, respectively, `OT-BERT(w/)' and `TOT-CAKT(w/)' are results with pretraining condition. From this table, we can observe that linguistic knowledge significantly enhances the ASR performance (not in a conventional way like LM rescoring). From the results in `OT-BERT(w/o)' and `TOT-CAKT(w/o)', both the methods in \cite{ASRU2023Lu} and the proposed cross-modal learning could efficiently transfer linguistic knowledge in acoustic encoding, yielding promising results. Moreover, when comparing results in `OT-BERT(w/o)' and `TOT-CAKT(w/o)', a significant performance improvement was observed when all models with cross-modal learning were trained from scratch (without pretraining condition). Furthermore, the performance of our proposed TOT-CAKT, even without pretraining could reach a level comparable to the original OT-based method, which requires a time-consuming pretraining stage (as in `OT-BERT(w/)). Finally, we further examined that when pretraining was utilized before cross-modal learning, the improvement of our proposed method (as shown in `TOT-CAKT(w/)' in table \ref{tab1}) compared to the `OT-BERT(w/)' in \cite{ASRU2023Lu} was reduced. This suggests that the pretraining stage could implicitly provide temporal order information for cross-modal feature alignment and knowledge transfer. In comparison, our proposed method explicitly incorporates temporal order information in the mathematical modeling with flexible parameters for control in experiments. Based on our formulation, our future work will focus on finding optimal temporal order parameter settings.   
\subsection{Visualization of transport coupling}
\begin{figure}[tb]
	%\vspace{3mm}
	\centering
	\includegraphics[width=8cm, height=5.0cm]{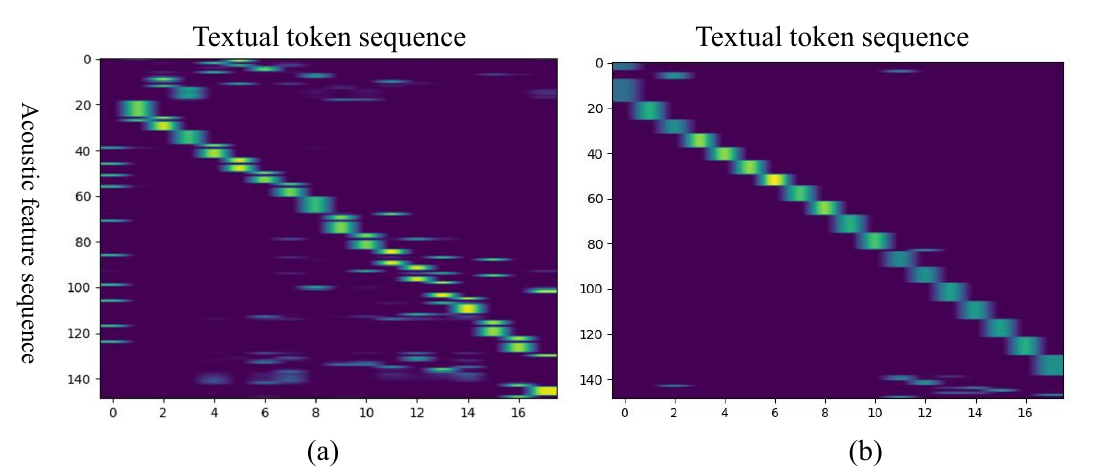}	
	\caption{Transport coupling matrix without temporal order constraint (a); with temporal order constraint (b). (Both of these cross-modal learning methods started from scratch without acoustic model pretraining.)}
	\vspace{-4mm}		
	\label{fig:coupling}
\end{figure}

In the proposed TOT-CAKT, the coupled pairs between the acoustic and linguistic feature sequences are explicitly designed to correspond to their temporal coherence, i.e., acoustic segments should match well with their linguistic tokens sequentially. Two examples of the coupling matrices are shown in Fig. \ref{fig:coupling}. In Fig. \ref{fig:coupling}-a, the coupling matrix is learned based on OT without temporal order constraint, and Fig. \ref{fig:coupling}-b is the coupling matrix learned with temporal order constraint. 
From this figure, it is evident that clear temporal correspondences exist between the acoustic feature sequence and linguistic token sequence in both transport couplings. Moreover, several positions with incorrect couplings in Fig. \ref{fig:coupling}-a were corrected in Fig. \ref{fig:coupling}-b by our proposed method, which explicitly adds a temporal order constraint.  
\section{Conclusion}
\label{sec:conclusion}
Acoustic and linguistic features belonging to different modalities require feature alignment as a crucial step in transferring linguistic knowledge from a PLM to acoustic encoding. In this paper, we proposed a novel TOT-CAKT. In the TOT-CAKT, a transfer coupling or mapping preserving temporal order information between acoustic sequence and linguistic sequence was first estimated. Subsequently, acoustic features were mapped to the linguistic space based on the transport coupling, allowing direct comparison of the mapped features to match the information encoded in the PLM. Our ASR experimental results confirmed the effectiveness of the proposed TOT-CAKT. Additionally, based on the visualization of transport coupling, we verified that TOT could eliminate unreasonable matches between the acoustic and linguistic sequences.   

In the proposed TOT-CAKT, several hyper-parameters involved in model learning are challenging to control. Additionally, some of these hyper-parameters are sensitive in implementation and may lead to stability problems in the Sinkhorn algorithm. In the current paper, the combined effects of those hyper-parameters have not been clearly explored, and a clear understanding of their combined rules has not been established, potentially hindering the identification of optimal solutions for improving ASR performance. Figuring out a set of optimal hyper-parameters in the proposed TOT-CAKT remains as our future work.

% References should be produced using the bibtex program from suitable
% BiBTeX files (here: strings, refs, manuals). The IEEEbib.bst bibliography
% style file from IEEE produces unsorted bibliography list.
% -------------------------------------------------------------------------
%\bibliographystyle{IEEEbib}
%\bibliography{strings,refs}

\bibliographystyle{IEEEtran}

\end{document}